\documentclass[aps,preprint, superscriptaddress,nofootinbib]{revtex4}
\usepackage{amsmath}
\usepackage{graphicx}
\usepackage{subfigure}
\usepackage{amssymb}
\usepackage{xcolor}
\usepackage{multirow}
\usepackage{cancel}
\usepackage{color}
\usepackage{epsfig}
\usepackage{ulem}
\usepackage{listings}
\usepackage{xcolor}
\usepackage{float}
\usepackage{slashed}

\usepackage{tikz}
\usepackage{pgffor}							

\usepackage[colorlinks,citecolor=blue]{hyperref}
\usepackage{amsmath}
\usepackage{wrapfig}

\newcommand{\be}{\begin{equation}}
\newcommand{\ee}{\end{equation}}
\newcommand{\beq}{\begin{equation}}
\newcommand{\eeq}{\end{equation}}
\newcommand{\bea}{\begin{eqnarray}}
\newcommand{\eea}{\end{eqnarray}}
\newcommand{\besp}{\begin{equation}\begin{split}}
\newcommand{\eesp}{\end{split}\end{equation}}

\newcommand{\slad}{\partial \!\!\!/}

\newcommand{\GeV}{\text{GeV}}
\newcommand{\MeV}{\text{MeV}}

\newcommand{\Dfbd}{\mathord{\buildrel{\lower3pt\hbox{$\scriptscriptstyle\leftrightarrow$}}\over {D}_{\mu}}}

\def\mL{\mathcal{L}}

\def\mO{\mathcal{O}}

\def\0{\textbf{0}}
\def\1{\textbf{1}}
\def\2{\textbf{2}}
\def\3{\textbf{3}}
\def\4{\textbf{4}}
\def\5{\textbf{5}}
\def\6{\textbf{6}}
\def\7{\textbf{7}}
\def\8{\textbf{8}}
\def\9{\textbf{9}}

\begin{document}

\title{A  hidden  self-interacting  dark matter sector with first order cosmological phase transition and gravitational wave}

\author{Wenyu Wang}
\email{wywang@bjut.edu.cn}
\affiliation{Faculty of Science, Beijing University of Technology, Beijing 100124, P. R. China}
\author{Wu-Long Xu}
\email{wlxu@emails.bjut.edu.cn}
\affiliation{Faculty of Science, Beijing University of Technology, Beijing 100124, P. R. China}
\author{Jin Min Yang}
\email{jmyang@itp.ac.cn}
\affiliation{CAS Key Laboratory of Theoretical Physics, 
	Institute of Theoretical Physics, 
	Chinese Academy of Sciences, Beijing 100190, P. R. China}
\affiliation{ School of Physical Sciences, University of Chinese Academy of Sciences, Beijing 100049, P. R. China}

\begin{abstract}
A dark scalar mediator can easily realize the self-interacting dark matter scenario and satisfy the constraint of the  relic density of the dark matter. When the hidden sector is highly decoupled from the visible sector, the constraints from direct and indirect detections of dark matter are rather relaxed.  The gravitational waves produced by the first order phase transition resulted from this dark scalar mediator  will be  an important signature to probe such a  dark sector.  
In this work a generic quartic finite-temperature potential is used to induce a strong first order phase transition. A joint analysis of the self-interacting dark matter,  the relic density of the dark matter and the first order phase transition shows that the mass range of the dark scalar is about  $(4\times 10^{-4} \sim 3)~\rm GeV$.  For the dark matter,  when the temperature ratio $\xi$ between the hidden sector and the visible sector is larger than 0.1, its mass range is found to be $(10~ \rm MeV\sim 10~ \rm GeV)$. The produced gravitational waves are found to have a peak frequency of $(10^{-6}\sim 10^{-3}) ~\rm Hz$ for a temperature  ratio $0.1<\xi<1$, which may be detectable in future measurements.
\end{abstract}

\maketitle

\tableofcontents

\section{Introduction}
The weakly interacting massive particle (WIMP) is the most popular dark matter candidate because its mass is around the electroweak scale and its thermal freeze-out naturally meets the observed relic density. 
This is the so-called WIMP miracle~\cite{Bertone:2016nfn,Lee:1977ua}. 
However, the recent direct detection of dark matter stringently limited the WIMP space, pointing to scenarios beyond the WIMP paradigm. On the other hand, despite of great successes achieved by the standard model of cosmology, i.e., the $\Lambda \rm CDM$ with cold dark matter as the dominant matter component for the evolution of the universe in the large scales, several problems seemingly appeared on the small cosmological scales, such as the core–cusp problem, the diversity problem and the too-big-to-fail problem~\cite{Tollerud:2014zha,Garrison-Kimmel:2014vqa,Navarro:1996gj,Moore:1999gc,Oman:2015xda}.  
From the side of particle physics, dark matter feebly interacts with the standard model (SM) particles. This suggests that dark matter may be a part of some hidden sector which almost decoupled from the visible sector at some temperature like the post-inflation reheating period \cite{Feng:2008mu, Berlin:2016gtr} in the early universe.  If so, the dark matter then froze out from the hidden sector at another specific low temperature with the evolution of the universe. 
After that, as the universe continued to cool down, the SM particles formed galaxies and the dark matter exists in the universe in the form of halos.  

Note that in the above-mentioned hidden sector dark matter scenario, the dark matter can have
self-interaction via exchanging a dark mediator. Such self-interacting dark matter (SIDM), unlike the collisionless dark matter in the $\Lambda \rm CDM$, can have elastic scattering between themselves and hence can solve those small-scale problems via the velocity dependence of the  self-interacting cross section per unit mass $\sigma /m$ which  is about $0.1-10~ \rm  cm^2/g$ in different small-scale structures \cite{Tulin:2017ara, Tulin:2013teo}. So in this scenario at least two new particles exist in the hidden sector, i.e., the dark matter particle and the dark mediator (scalar or vector) ~\cite{Bringmann:2020mgx, Dery:2019jwf}. Note that in this case the stability of the dark mediator should also be carefully checked when considering the relic density of dark matter ~\cite{Duerr:2018mbd,Kahlhoefer:2017umn}.  On the  one hand, if the decoupling between hidden sector and visible sector is incomplete, the thermal equilibrium may be maintained via the decay of this mediator into the SM particles after the freeze-out of the dark matter.  The coupling  strength of the portal between hidden sector and visible sector will be severely constrained by the dark matter direct detection.  Anyway, the life-time of this mediator will be limited so that it does not spoil the light species abundances predicted by the Big Bang Nucleosynthesis (BBN)~\cite{Kaplinghat:2013yxa}. 
On the other hand, if hidden sector and visible sector are highly decoupled, 
it will be much favored because the constraints of
dark matter direct detection can be relaxed and both the SIDM paradigm  
and the demanded dark matter relic density can be  realized easily. However, if the mediator is stable, this particle still may dominate the energy density 
of the early universe in the non-relativistic case ~\cite{Berlin:2016gtr}. 
In other words, the decay of this mediator into the SM particle are 
strictly constrained (except that there exist other dark particles which can decay 
into the SM particles to relax the dark matter direct detection limits)  
and should  be further constrained by the Cosmic Microwave Background (CMB) and BBN \cite{Duch:2019vjg}.  
Therefore, the dark matter direct detection, together with CMB and BBN, will stringently constrain the coupling strength of the portal between the hidden sector and visible sector and also constrain the life-time of the portal. 

\begin{figure}[h]
	\centering
	\includegraphics[width=12cm]{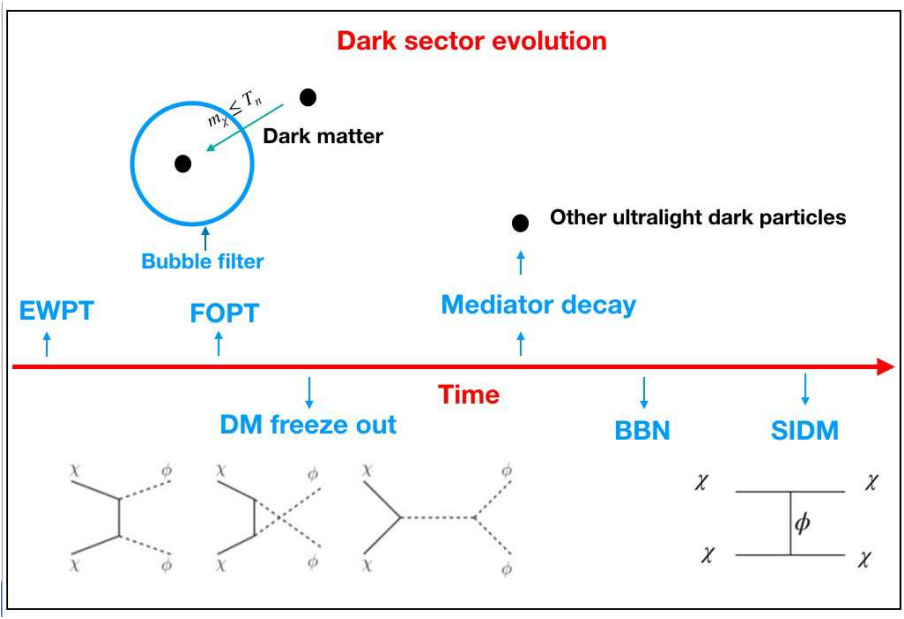}
	\caption{The order of FOPT, freeze out of DM, mediator decay and SIDM in the dark sector evolution history in our work. }
	\label{fig11}
\end{figure}
In view of a wide variety in the hidden sector, we consider a simple hidden sector with a Dirac fermion dark matter and a scalar to mediate self-interaction of dark matter. We assume the hidden sector to  
 highly decouple from the visible sector.     
In this scenario, the dark sector can have gravitational waves (GWs) 
produced by the first order phase transition (FOPT) in the early universe
\cite{Fairbairn:2019xog,Breitbach:2018ddu,Dent:2022bcd,Marfatia:2020bcs,Han:2020ekm,Athron:2019teq,Kawana:2022lba,Bian:2021vmi,Bian:2019kmg,Hall:2019rld, Hall:2019ank,Azatov:2021ifm, Azatov:2022tii,Konoplich:1999qq,Dymnikova:2000dy}. 
Then various constraints on this hidden SIDM, such as from the relic density,  will be studied in this work. If a strong FOPT happens,  the hidden sector physics can be accessible through the detection of the GWs in the future.  The time order of the FOPT, the freeze-out of DM, the mediator decay and the SIDM in the dark sector evolution in our model is shown in Fig. \ref{fig11}. Note that for our hidden SIDM sector, although we only present the dark matter and the dark scalar,  this dark scalar must decay into other ultra-light dark particles (say some ultra-light dark fermions) after the dark matter freeze out, which will not affect the dark matter annihilation process and the dark matter relic density.  The abundances of the light species in the BBN stage  will not be spoiled by the dark particles.  This means that when the hidden sector is highly decoupled from the visible sector,  other light dark  particles must exist unless the dark mediator can decay into some SM particles via the freeze-in mechanism after the dark matter freeze out from the hidden sector.  In this work, our focus is the dark matter, which can penetrate the  bubble filter of FOPT  and satisfy the constraint of relic density, and the dark scalar which can realize the SIDM paradigm and FOPT in the hidden sector. 
This work is organized as the follows.  A benchmark model with cosmological phase transition and gravitational wave is described in Sec. II.   The study of the SIDM in this model is given in Sec. III.   
The  numerical results  are presented in Sec. VI and the conclusion is given in Section VII.

\section{A simple model with FOPT in hidden sector }\label{sec1}	
\subsection{A benchmark model}\label{subsec21}	
If a hidden sector exists,  the dark sector can be highly decoupled from the visible sector  
after the inflation or since the third stage of the reheating  at a specific dark temperature.  To be simplest,  we assume the existence of a dark Dirac fermion $\chi$ as the dark matter and a dark scalar particle $\phi$ with the dark Yukawa interaction \cite{Marfatia:2021twj}. 
The Lagrangian is 
\beq  \label{lag}
\mL\supset\bar{\chi}i\slad\chi-g_{\chi}\phi\bar{\chi}\chi-V_{eff}(\phi,T),
\eeq
where $\chi$ has a global $U(1)$ symmetry and   $V_{\rm eff}(\phi,T)$ 
is the finite-temperature effective potential of  the field  $\phi$.    
When temperature decreases to a critical temperature $T_c$,  
the universe abruptly traverses from a meta-stable state to another ground state. 
This means that the scalar field $\phi$ is quantum-tunneling from the false vacuum $\langle\phi\rangle=0$ to the true vacuum $\langle\phi\rangle=v_{\phi}$.  
As studied in Ref.~\cite{Hong:2020est}, the bubble wall plays a role of a filter during the FOPT.  According to the energy conservation,  if the mass of $\chi$ in the true vacuum is smaller than its kinetic energy in the false vacuum,  all particle can penetrate the bubble wall safely, namely  $m_{\chi} \simeq g_{\chi}v_{\phi}\leq T_n$ (note $T_n$ is the following nucleation temperature $T_{hn}$).   Due to no residual dark matter in the false vacuum, the primordial black holes, the Fermi-balls, the Q-balls, or the thermal balls will not be formed \cite{Gross:2021qgx,Huang:2022him,Baker:2021nyl,Baker:2021sno,Huang:2017kzu}. Subsequently, we consider the case that the freeze-out dark matter in this hidden sector  will account for all the required relic density. 
To solve the small scale problems by this hidden self-interacting dark matter, 
the mass of scalar mediator $\phi$ is above $\MeV$, as found in our following study,  which is larger than the photo temperature at neutrino decoupling \cite{Breitbach:2018ddu}.  
As a result, the cosmological constraints such as CMB and BBN can be avoided 
during the FOPT in our analysis.

\subsection{Cosmological phase transition and gravitational waves}\label{subsec22}
Superficially, our SIDM model is kind of 'effective' description which just contains a dark Dirac fermion $\chi$ as the dark matter and a dark scalar particle $\phi$.  
In a complete theory, the particles in the hidden sector could be similar to the visible sector,
i.e., with additional dark particles like the right-handed neutrinos~\cite{DiBari:2021dri} 
and the dark complex scalars in various supersymmetric models \cite{Wang:2022lxn,Huang:2014ifa}.     Meanwhile, after the dark matter freeze out, the dark mediator must decay into other light dark particles. The light species abundances in the BBN stage  will not be spoiled by the dark particles. 
Since our model is only an 'effective' description,  we choose a generic form of the quartic finite-temperature potential for the dark scalar \cite{Adams:1993zs, Kehayias:2009tn}
\beq \label{potential}
V_{\rm eff}(\phi,T_h)=D(T_h^{2}-T_{h0}^{2})\phi^{2}-(AT_h+C)\phi^{3}+\frac{\lambda}{4}\phi^{4},
\eeq 
where $D$, $A$ and $\lambda$ are dimensionless parameters,  $C$ provides the cubic term at zero temperature and  $T_h$ is the temperature of the hidden sector.  
Simply, the minimal value of the potential is located at 
\beq \label{mini}
v_{0\pm}=\frac{3C\pm\sqrt{9C^{2}+8\lambda DT_{h0}^{2}}}{2\lambda},
\eeq 
Note that here is the zero temperature case, $T_h=0$. The mass of the scalar particle at zero temperature is obtained from 
the second derivative of $V(\phi, 0)$ with respect to $\phi$
\beq \label{mass}
m_{\phi}=\frac{d^{2}V_{\rm eff}(\phi,0)}{d\phi^{2}}\Big|_{\phi=v_{0+}}=4DT_{h0}^{2}+3Cv_{0+} \, .
\eeq 
The critical temperature is obtained from $V(0,T_{c})=V(v(T_{c}),T_{c})$ 
\beq \label{crit}
T_{hc}=\frac{-CA-\sqrt{D\lambda(C^{2}-(A^{2}-D\lambda)T_{h0}^{2})}}{A^{2}-D\lambda}.
\eeq 
The corresponding scalar field value will be
\begin{eqnarray}
 \phi_{T_{hc}}=\frac{3C+3AT_h+\sqrt{(-3C-3AT_h)^{2}-4(2dT_h^{2}-2dT_{h0}^{2})\lambda}}{2\lambda}. 
\end{eqnarray}
 Note that   $A^2-D\lambda<0$ is required for a real critical temperature $T_{hc}$.

During the FOPT process, the decay probability per unit time per unit volume is 
 \be \label{Gamma}
 \Gamma\sim T_h^4\left(\frac{S_3(T_h)}{2\pi T_h}\right)^{3/2}e^{-S_3(T_h)/T_h},
 \ee
where  the Euclidean action $S_3(T)$ is written as
\be\label{S_3}
S_3=\int_0^\infty 4\pi r^2dr\left[\frac12\left(\frac{d\phi}{dr}\right)^2+V_{\rm eff}(\phi,T_h)\right].
\ee
To evaluate $S_3$, the $\mO(3)$ symmetric equation of motion needs to be solved, namely,
\be
\frac{d^2\phi}{dr^2}+\frac{2}{r}\frac{d\phi}{dr}=\frac{\partial }{\partial\phi}V_{\rm eff}(\phi,T_h),\quad
\lim_{r\to\infty}\phi=0,\quad\frac{d\phi}{dr}\Big|_{r=0}=0.
\ee
For a three-dimensional Euclidean action, it can be derived from a quartic potential
\beq\label{newpotential}
V_{\rm eff}(\phi, T_h)\simeq  \bar{\lambda}\phi^{4}-a\phi^{3}+b\phi^{2},
\eeq 
These parameters are given as 
\begin{eqnarray}
&&\bar{\lambda}=\lambda/4,\\
&&a=AT_h+C,\\
&&b=D(T^2_h-T^2_{h0}).
\end{eqnarray}

The corresponding semi-analytic approximate  $S_3$ is given by 
\beq \label{neqS3}
S_{3}(T_h)=\frac{\pi a}{\bar{\lambda}^{\frac{3}{2}}}\frac{8\sqrt{2}}{81}(2-\delta)^{-2}\sqrt{\frac{\delta}{2}}(\beta_{1}\delta+\beta_{2}\delta^{2}+\beta_{3}\delta^{3}),
\eeq 
with $\delta=8\bar{\lambda}b/a^2$, $\beta_1=8.2938$, $\beta_2=-5.5330$  
and $\beta_3=0.8180$~\cite{Adams:1993zs}.

The probability for a bubble to nucleate inside a Hubble volume is
\be
N(T_h)=\int_{T_{h}}^{T_{hc}}\frac{dT'_h}{T'_h}\frac{\Gamma(T'_h)}{H^4(T'_h)},
\ee
where $H(T_h)$ is the Hubble constant.   When $N(T_{hn})\sim1$, the solved temperature is
called the nucleation temperature $T_{hn}$.  For a FOPT around  
$\mO(100\GeV)$, the nucleation temperature can be approximately taken as ~\cite{Quiros:1999jp}
\be\label{S3T}
S_3(T_{hn})/T_{hn}\simeq 140,
\ee
which is called as the electroweak phase transition and , this value  can be derived from
the following Eq.(\ref{OrigS3T}).
In fact, the nucleation temperature is not necessarily at 
the electroweak scale, whose criterion can be expressed as \cite{Imtiaz:2018dfn}
\be\label{OrigS3T}
\frac{S_3}{T_{hn}}\simeq \ln [\frac{1}{4}(\frac{90}{8\pi^3 g_{\rm eff}})^2]+4 \ln [\frac{M_{pl}}{T_{hn}}].
\ee
Here, $T_{hn}$ on the right-hand side is taken place by  
$T_{hc}$  approximately due to  the logarithmic function.

In the FOPT, the  parameter $\alpha$ denotes the strength of the phase transition, which is read as 
\beq \label{alpha}
\alpha=\frac{(1-T\frac{\partial}{\partial T_h})\Delta V\big|_{T_{h*}}}{\rho_r} ,
\eeq 
where $\Delta V=V_{\rm eff}(0,T_h)-V_{\rm eff}(v_{\phi}(T_h),T_h)$ 
and  the radiation energy density is $\rho(T_h)=\frac{\pi^2}{30}g_*T^4_{SM}$ 
with $g_*=g_{*SM}+g_{*D}(\frac{T_h}{T_{SM}})^4$ .  We take $g_{*D}=4.5$ 
at all relevant time and $g_{*SM}$ can be found in ~\cite{Husdal:2016haj}.  
Because the energy of FOPT we are interested in is above $\MeV$,  
the temperature ratio between hidden sector and visible sector is not constrained by the BBN or CMB.  
Another parameter is the inverse duration of the phase transition which can be written as
\beq \label{beta}
\beta=\frac{\dot{\Gamma}}{\Gamma}\simeq-\frac{d(\frac{S_{3}}{T_h})}{dt}\big|_{t=t_{h*}}.
\eeq 
In the  GW calculation, it is expressed as 
\beq\label{betaH}
\frac{\beta}{H_{*}}\simeq T_{h*}\frac{d(\frac{S_{3}}{T_h})}{dT_h}\big|_{T_{h*}}
\eeq 
Generally, a bigger  $\alpha$ and a smaller $\beta$ imply a stronger FOPT.

Then the main way to generate stochastic GWs in the FOPT  includes bubble collision, sound waves and turbulence of the magneto-hydrodynamics (MHD) in the particle bath.  Due to the friction the motion of bubble walls will be significantly dampened in the plasma-wall system. So the wall will reach a terminal velocity $v_b$ at a very short time.  Therefore, the bubble collision contribution is negligible. Note that our work focuses on  CPT and GW, so  we consider that when bubble walls collide, the thickness of walls will be very thin. Thus  this process does not form a PBH \cite{Jung:2021mku}. Certainly, under certain conditions, the closed walls may eventually collapse into primordial black holes. However, the abundance of these primordial black holes should not be very large so that it does not contradict the observations \cite{Deng:2016vzb,Belotsky:2018wph,Rubin:2001yw,Carr:2019bel}.   On the flip side, most of energy is pumped into the fluid shells surrounding the wall~\cite{Ellis:2018mja, Wang:2020jrd}.   Finally the significant contribution source of the GWs is from the sound waves while the contribution of the MHD turbulence is a sub-leading source for the GWs.
The GW spectrum is 
\be\label{GW}
\Omega_{\rm GW}(f)=\frac{1}{\rho_c}\frac{\rho_{\rm GW}}{d\ln f},
\ee
where $f$ is the frequency, $\rho_{\rm GW}$ is the GW energy density produced during FOPT  
and $\rho_c$ is the critical energy density of the present universe. 
The total  GW spectrum is written as 
\be\label{GWsignal}
\Omega_{\rm GW}h^{2}\simeq\Omega_{\rm sw}h^{2}+\Omega_{\rm turb}h^{2},
\ee
where the sound wave contribution is written as 
\begin{eqnarray}
&&h^{2}\Omega_{\rm sw}(f)=2.65\times10^{-6}(H_n\tau_{\rm sw})\left(\frac{H_{n}}{\beta}\right)\left(\frac{\kappa_{v}\alpha}{1+\alpha}\right)^{2}\left(\frac{100}{g_{*}}\right)^{\frac{1}{3}}v_bS_{\rm sw}(f),\\
&&S_{\rm sw}(f)=\left(\frac{f}{f_{\rm sw}}\right)^{3}\left(\frac{7}{4+3(\frac{f}{f_{\rm sw}})^{2}}\right)^{\frac{7}{2}},\\
&&f_{\rm sw}=1.9\times10^{-2}~{\rm mHz}\frac{1}{v_b}\left(\frac{\beta}{H_{n}}\right)\left(\frac{T_{hn}}{100~{\rm GeV}}\right)\left(\frac{g_{*}}{100}\right)^{\frac{1}{6}},
\end{eqnarray}
with $\tau_{\rm sw}$ being the duration of the sound wave source  and  $\kappa_v$ being the ratio of the bulk kinetic energy to the vacuum energy ~\cite{Espinosa:2010hh, Wang:2020jrd}.
The turbulence contribution is written as
\begin{eqnarray}\label{turbulence}
&&h^{2}\Omega_{{\rm turb}}(f)=3.35\times10^{-4}(\frac{H_{n}}{\beta})(\frac{\kappa_{\rm turb}\alpha}{1+\alpha})^{2}(\frac{100}{g_{*}})^{\frac{1}{3}}v_{w}S_{\rm sturb}(f)\\
&&S_{\rm turb}(f)=\frac{(\frac{f}{f_{\rm turb}})^{3}}{[1+(\frac{f}{f_{\rm turb}})]^{\frac{11}{3}}(1+\frac{8\pi f}{h_{p}})}\\
&&f_{\rm sw}=2.7\times10^{-2}{\rm mHz}\frac{1}{v_{w}}(\frac{\beta}{H_{n}})(\frac{T_{hn}}{100Gev})(\frac{g_{*}}{100})^{\frac{1}{6}}\\
&&\kappa_{\rm turb}=\epsilon\kappa_{v}.
\end{eqnarray}
Here $\epsilon$  represents  the fraction of bulk motion which 
is turbulent and $H_{n}$ is the Hubble parameter at $T_{hn}$.

\section{Dark matter relic density and self-interaction}\label{sec2}
\subsection{Dark matter relic density}\label{subsec31}

After the dark matter safely penetrates the bubble walls, 
the FOPT can be considered as decoupling from the freeze-out process of the dark 
matter~\cite{Chao:2020adk}. In contrast with the usual WIMP freeze-out mechanism \cite{Liu:2013vha},  
the Boltzmann equation needs to be modified in some aspects, e.g., the equilibrium density $n_{\rm eq}$ and the thermal average $\langle\sigma v\rangle$  
are evaluated at the dark temperature $T_h$ rather than the SM temperature $T$ and 
the Hubble parameter must be included in  the energy of the dark sector \cite{Bringmann:2020mgx}. 
Next we examine the dark matter relic density when the hidden sector 
and the visible sector are at different temperature. Here the temperature ratio 
is defined as $\xi=T_h/T$. The cosmological evolution of the dark matter
is determined by the following Boltzmann equation which takes a similar form as in the WIMP paradigm: 
\beq \label{Bolzeq}
\frac{dn_{\chi}}{dt}+3Hn_{\chi}=\langle\sigma v\rangle((n_{\chi}^{\rm eq})^{2}-n_{\chi}^{2}),
\eeq 
where  the equilibrium number density is in the non-relativistic limit
\beq\label{end}
n^{\rm eq}_{\chi}=g_{\chi}(\frac{\xi m_{\chi}T}{2\pi})^{\frac{3}{2}}e^{-\frac{m_{\chi}}{\xi T}}.
\eeq 
The Hubble parameter during radiation domination is given by  
\beq \label{Hp}
H^2 =\frac{8\pi^3}{90}g_{\rm eff}M^{-2}_{\rm PI}T^4,
\eeq
where the total effective relativistic energy degree-of-freedom (d.o.f) 
$g_{\rm eff}=g_{\rm SM}+(\sum_bg_b+\frac{7}{8}\sum_fg_f)\xi^4$ with $g_b$ and  $g_f$ being respectively the intrinsic d.o.f of bosons (b) and fermions (f) in the dark sector.  In our work,  we take $g_b=1$ and $g_f=3.5$ for numerical calculations.

\begin{figure}[h]
	\centering
	\includegraphics[width=12cm]{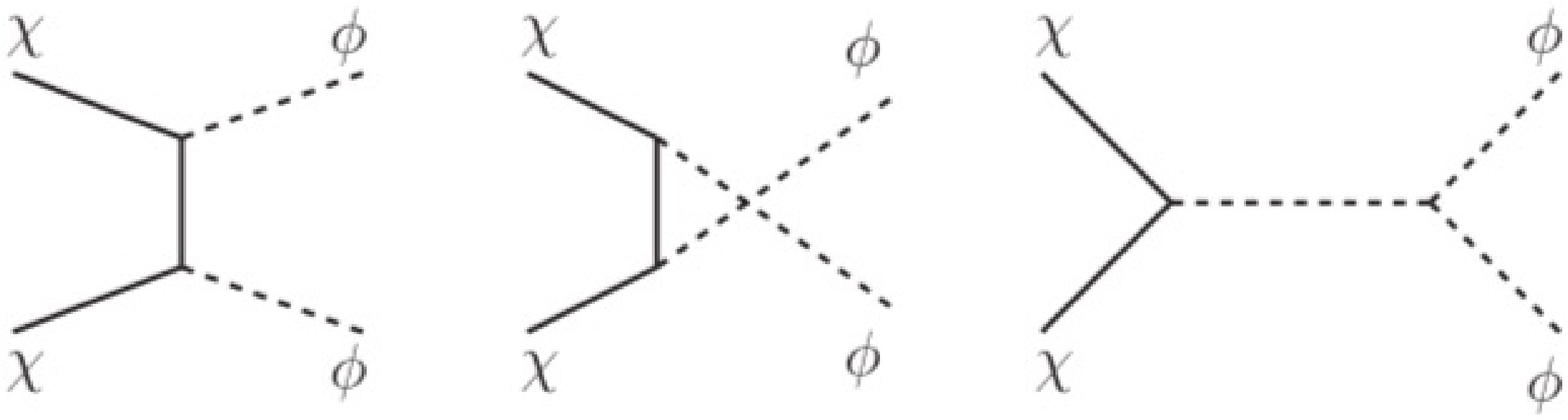}
	\caption{The Feynman pictures from three annihilation processes $\chi \chi \rightarrow \phi\phi$}
	\label{fig1}
\end{figure}
In the non-relativistic regime,  according to the Feynman diagrams in 
FIG.~\ref{fig1}, the approximate thermal average pair annihilation 
cross section of the dark matter is 
\beq \label{ance}
\langle\sigma v\rangle=a+b\langle v^2\rangle=a+6\xi b\frac{T}{m_{\chi}},
\eeq 
where $a$ is the $s$-wave cross section and $b$ is the $p$-wave cross section.
In this scenario, $\langle\sigma v\rangle$ is given by ~\cite{Shtabovenko:2016sxi,Shtabovenko:2020gxv}
\beq \label{app}
\begin{split}
\langle\sigma v\rangle=&\frac{mv^{2}}{96\pi(m_{\phi}^{2}-2m_{\chi}^{2})^{4}(m_{\chi}m_{\phi}^{2}-4m_{\chi}^{3})^{2}}g_{\chi}^{2}\sqrt{(m_{\chi}-m_{\phi})(m_{\chi}+m_{\phi})}\\
&\times (3C^{2}(m_{\phi}^{2}-2m_{\chi}^{2})^{4}+4Cg_{\chi}m_{\chi}(28m_{\chi}^{4}-11m_{\chi}^{2}m_{\phi}^{2}+m_{\phi}^{4})(m_{\phi}^{2}-2m_{\chi}^{2})^{2}\\
&+8g_{\chi}^{2}m_{\chi}^{2}(m_{\phi}^{2}-4m_{\chi}^{2})^{2}(9m_{\chi}^{4}-4m_{\chi}^{2}m_{\phi}^{2}+m_{\phi}^{4})).
\end{split}
\eeq 
This  implies that the annihilation is dominated by $p$-wave process.

As studied in ~\cite{Chacko:2015noa}, the freeze out temperature can be evaluated by
\beq \label{xf}
x_{f}=\xi\log[\frac{c(c+2)}{4\pi^{3}}\sqrt{\frac{45}{2}}\frac{g_{\chi}}{\sqrt{g_{*\rm eff}}}m_{\chi}M_{\rm pl}\frac{\xi^{\frac{5}{2}}\langle\sigma v\rangle}{\sqrt{x_{f}}}]
\eeq  
where $x_f=m_{\chi}/T$, and $c=\sqrt{2}-1$ is for $s$-wave annihilation 
and $c=\sqrt{3}-1$ is for $p$-wave annihilation. 
Finally, the  dark matter relic density is given by
\bea \label{relic}
\Omega_{\rm DM}h^2=m_{\chi}s_0Y_{\infty}\rho^{-1}_c ,
\eea 
in which 
\bea \label{relicxxx}
Y_{\infty}=\frac{x_{f}}{\sqrt{\frac{\pi}{45}}\frac{g_{*}}{g_{\rm eff}^{1/2}}M_{\rm pl}m_{\chi}(a+3\xi b/x_f)},
\eea 
and  $s_0$ and  $\rho_c$  are the entropy density and critical density at present time, respectively.

\subsection{Dark matter self-interaction }\label{subsec32}
As mentioned above, an appropriate self-interaction between the dark matter particles can give the required scattering cross section per unit mass $\sigma/m$ \cite{Kim:2022cpu,Kim:2021bmx,Bringmann:2016din,Chu:2018fzy}. Especially, the interaction through a light mediator may give 
a specific required velocity-dependence for different small scale objects. 
Note that another source of velocity-dependence can be obtained by including 
the finite-size effect~\cite{Chu:2018faw,Wang:2021tjf}.  
The transfer cross section is written as  
\begin{equation}\label{sigt}
\sigma_{\rm T} \equiv \int d\Omega \, (1-\cos \theta)\frac{d\sigma}{d\Omega} .
\end{equation}
Generally, this cross section should be obtained by solving the Schr\"odinger 
equation numerically. 
In our work, the calculation method shown in \cite{Colquhoun:2020adl} is used. 
Two dimensionless parameters $\kappa$ and $\beta$ are used to delineate different 
regimes such as the Born ($2\beta\kappa^2 \ll 1$), the quantum ($\kappa\ll1$) 
and the semi-classical ($\kappa\geq1$). 
These two parameters are 
\beq \label{par}
\kappa = \frac{m_{\chi} v}{m_{\phi}},\quad\quad  \beta=\frac{2\alpha m_{\phi}}{m_{\chi}v^2 } .
\eeq 
The  analytic formulas of  the 
transfer cross section in the semi-classical regime ($\kappa\geq1$) for an attractive Yukawa potential are given by  ~\cite{Colquhoun:2020adl}
\begin{align}\label{sigTatt}
	\sigma_{T}^{\rm{att}} & = \frac{\pi}{m_\phi^2}  \times \begin{cases}
		2\beta^2 \zeta_{1/2}\left(\kappa, \beta\right) & \beta\leq0.2\\
		\hspace{7cm}\  & \ \\[-4.1mm]
		2\beta^2 \zeta_{1/2}\left(\kappa, \beta\right) e^{0.64(\beta - 0.2)} & 0.2 < \beta \leq 1\\
		\hspace{7cm}\  & \ \\[-4.1mm]
		4.7 \log(\beta + 0.82) & 1< \beta < 50 \\
		\hspace{7cm}\  & \ \\[-4.1mm]
		2\log\beta(\log\log \beta+1)& \beta\geq50,
	\end{cases}
\end{align}
where
\begin{align}\label{zeta}
\zeta_n(\kappa,\beta) & =\frac{\text{max}(n,\beta\kappa)^2 - n^2}{2 \kappa^2 \beta^2} + \eta\left(\frac{\text{max}(n,\beta\kappa)}{\kappa}\right)\,,\\
\eta(x) & = x^2 \left[ -K_1\left(x\right)^2 + K_0\left(x\right) K_2\left(x\right)\right]\,,
\end{align}
with $K_n$ being the modified Bessel function of the second kind.
In case of $\kappa<0.4$, the  Schr\"odinger equation can be solved 
analytically via the Hulth\'en potential since this quantum regime 
is dominated by $s$-wave scattering \cite{Tulin:2013teo}, 
\begin{equation}\label{Hulthenpotential}
\sigma_{T}^{\mbox{Hulth\'en}}=\frac{16\pi}{m_{\chi}^2v^2}\sin^2\delta_0,
\end{equation}
where the phase shift $\delta_0$ is
\begin{equation}\label{phaseshift}
\begin{split}
\delta_0&=\arg\left(i \frac{\Gamma(l_{+} + l_{-} -2)}{\Gamma(l_{+})\Gamma(l_{-})}\right),\\
l_{+}&=1+\frac{\kappa}{1.6}(i+i\sqrt{3.2\beta\pm 1}),\\
l_{-}&=1-\frac{\kappa}{1.6}(i+i\sqrt{3.2\beta\pm 1}).
\end{split}
\end{equation}
Here the signs $+(-)$ denotes repulsive (attractive).
When $\kappa$ is in the range   $(0.4, 1)$,   as shown in~\cite{Colquhoun:2020adl}, 
the interpolation function is used, namely,  
\begin{eqnarray}
  \sigma_{T}=(1-\kappa)/0.6 \sigma_{T}^{\mbox{Hulth\'en}}+(\kappa-0.4)/0.6 \sigma_{ T}^{\rm{rep(att)}}.
\end{eqnarray}
Thus,  the entire  interesting parameter space of SIDM can be  almost covered analytically. Using the Maxwell-Boltzmann distribution, we can get the 
the velocity-averaged transfer cross section  
\bea \label{vat}
&&\langle \sigma_{T}v \rangle=\int f(v)\sigma_T vdv\\
&&f(v)=\frac{32v^2e^{-4v^2/\pi\langle v\rangle^2}}{\pi^2\langle v\rangle^3}
\eea 
where $v$ is the relative velocity in the center-of-mass frame.

\section{Numerical results}\label{sec4}
In this section  we show the numerical results of the FOPT and the constraints
on SIDM including the attractive Yukawa potential  and dark matter relic density.
We will concentrate on the favored mass parameter space of the dark matter and 
the scalar particle. 

\begin{figure}[h]
	\centering
	 \includegraphics[width=5.25cm]{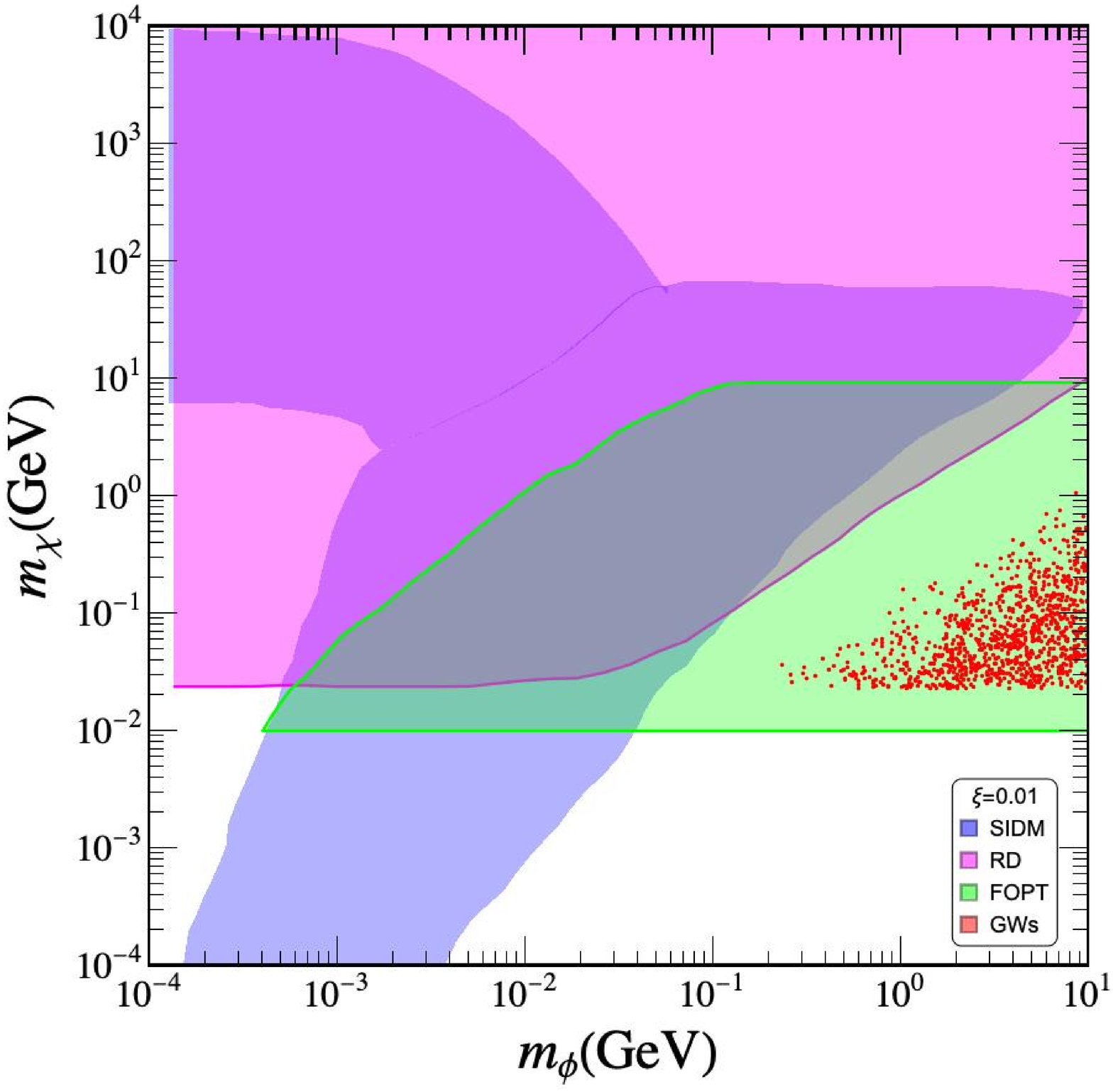}
	 \includegraphics[width=5.25cm]{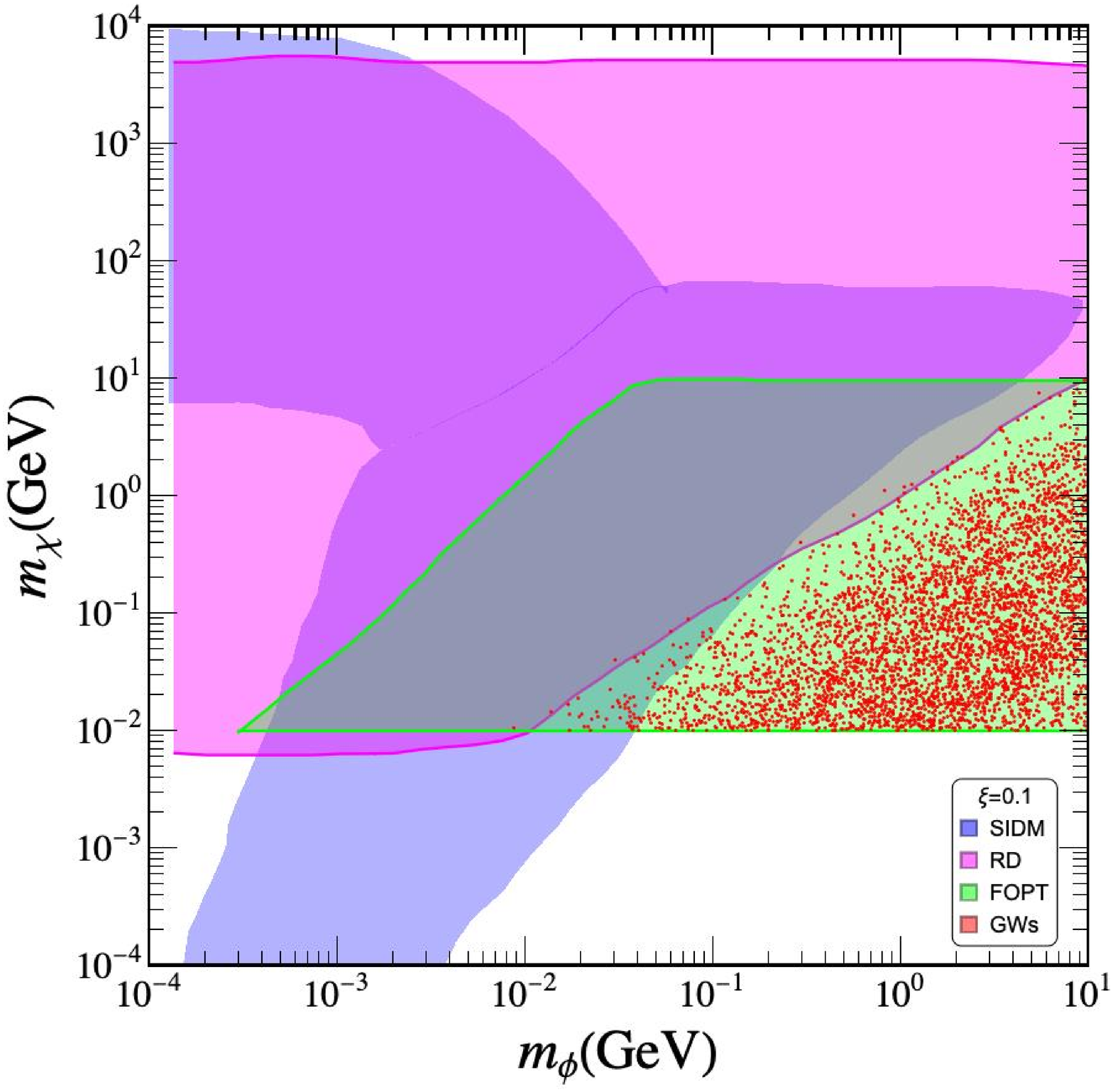}
	 \includegraphics[width=5.25cm]{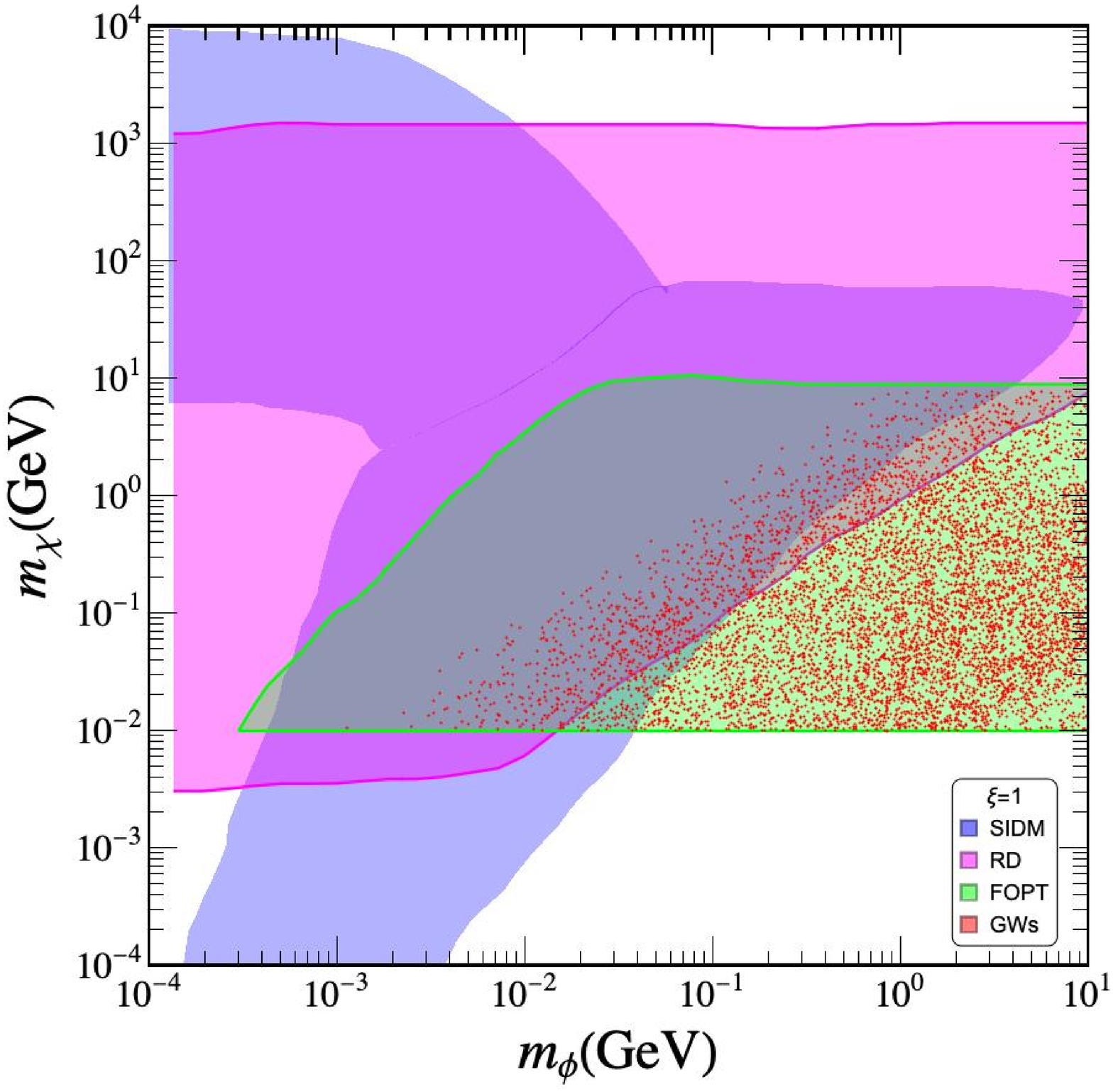}
	\caption{The parameter space of $m_{\phi}$ and $m_{\chi}$ for different  temperature ratio $\xi =0.01, 0.1, 1$. The shadow areas correspond to three constraints: SIDM (blue for attractive Yukawa potential), dark matter relic density (magenta) and FOPT (green). The red scatter points are for the produced GWs with power spectrum  $\Omega_{\rm GW}h^2>10^{-20}$.}
	\label{fig2}
\end{figure}
First, in the SIDM scenario for an attractive Yukawa potential, 
the input parameters include the masses of dark matter $m_{\chi}$ and     
dark scalar $m_{\phi}$ as well as the dark Yukawa coupling constant $g_{\chi}$. 
These parameters vary in the ranges 
\beq \label{pre}
10^{-4}~{\rm GeV}<m_{\chi}< 10^4~ {\rm GeV}, \quad  
10^{-4}~{\rm GeV}<m_{\phi}< 10^4 ~{\rm GeV},\quad  
10^{-3}<g_{\chi}< 1
\eeq
To solve the small-scale structure anomalies,  the value of $\sigma_T/m_{\chi}$ 
in the range $(0.1,10 ~\rm cm^2/g)$ is favored.  Then the favoured regions 
of the masses $m_{\chi}$ and $m_{\phi}$ are shown as the blue part in Fig.~\ref{fig2}.  It shows a significant mass split between those two dark particles. 
Unlike the results in Ref.~\cite{Tulin:2013teo} which gives an allowed 
parameter space of $m_{\phi}$ in ($1\sim 100$) MeV, here the allowed mass space is larger for the attractive SIDM case. The reason is that our model has a more completed self-interaction in the hidden sector. Another point should be noted is that the resonant effect is very obvious and the three panels of Fig.~\ref{fig2} imply that the up bound for $m_{\phi}$ is about 10  GeV.  For the input parameters during the FOPT, we choose 
\beq \label{ipp}
\begin{split}
&10^{-4}<A< 10^4, \quad  
10^{-4} <D< 10^4, \quad  
10^{-4}~{\rm GeV}<T_0< 10^4 ~{\rm GeV},\quad  \\
&
10^{-4}~{\rm GeV}<C< 1 ~{\rm GeV},\quad  
10^{-4}<\lambda< 1.
\end{split}
\eeq
The corresponding numerical results are shown in the green area in Fig. ~\ref{fig2}.
For the FOPT  the maximum  mass of  dark matter $m_{\chi} \simeq T_n$ is taken. For example,  for a benchmark point with $m_{\chi}=0.22322$ and $v_{\phi}=1.6682$, we have  $g_{\chi}=0.13381$.   
The different $\xi$  values almost have no influence for the selection of dark particle mass.  The combination of SIDM and FOPT requires $m_{\chi} $ in the range of $ (10^{-2} \sim 10)$   GeV.

Finally we show the constraints from the dark matter relic density $\Omega_{\rm DM}h^2 \in(0.11,0.13)$.  Now we have one  more coupling coefficient $C$, which denotes the cubic term in Eq.(\ref{potential}), and we set it in the range  of $(10^{-3}\sim 1)$ GeV.   
As the initial temperature ratio between those two sectors after the reheating 
is unknown, we simply take the temperature ratio  $\xi =0.01, 0.1, 1$  
for the calculation of the dark matter relic density.  
The numerical results are shown in the magenta region of  Fig.~\ref{fig2}, 
which indicate that the lowest bound for $m_{\chi}$ is higher than 
$3 \times 10^{-3}$ GeV when the hidden sector is colder than the visible sector.  
In all, the SIDM and the relic density of dark matter can be satisfied by the sub-GeV dark scalar and in this hidden sector the FOPT can happen. To achieve all these, the mass space is about $m_{\chi} \in (0.01\sim 10)$ GeV and $m_{\phi} \in (4\times 10^{-4}\sim 3)$ GeV when the hidden sector is colder than the visible sector, as shown by the gray (overlapped) region in Fig. \ref{fig2}. When the value of $\xi$ is smaller, the survived space of $m_{\chi}$ becomes narrower. Note that in our study the dark matter froze out after the FOPT so that its relic density is not diluted by the FOPT \cite{Xiao:2022oaq}. 

\begin{figure}[h]
	\centering
	\includegraphics[width=8cm]{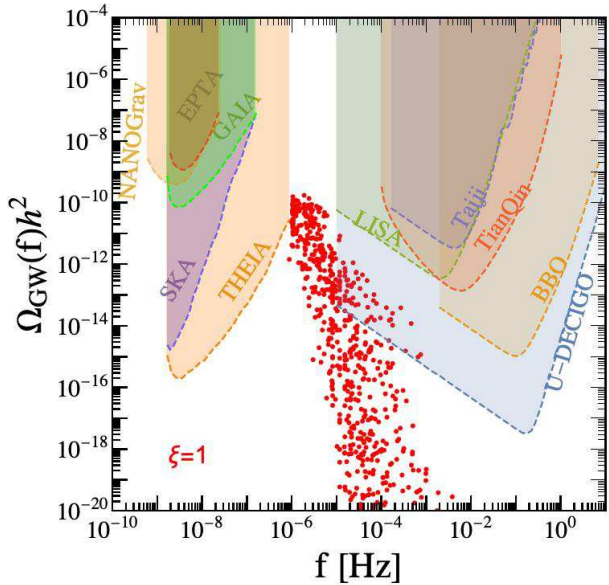}
	\caption{The GW power spectrum $\Omega_{\rm GW}(f)h^2$.  The red points denote the GWs produced by the FOPT of our model for a temperature ratio $\xi=1$. 
 The detectable regions of various GW detectors are shown as the shadow areas. 
}
	\label{fig3}
\end{figure}

In order to find out whether the GWs produced by FOPT can be detected,  
we calculate the GW power spectrum $\Omega_{\rm GW}h^2$ in the parameter space  $m_{\chi} \in (0.01\sim 10)$  GeV and  $m_{\phi} \in (4 \times 10^{-4}\sim 3)$  GeV which satisfy the FOPT.  
Then the corresponding GW power spectrum $\Omega_{\rm GW}h^2>10^{-20}$ is plotted as red points in Fig. \ref{fig2}.  We can see that the red points are
 excluded by the constraint of dark matter relic density for $\xi=0.01, 0.1$, 
indicating that the GW power spectrum cannot be detected by current detectors. This is because the latent energy is quite small in these cases. 
Instead,  for $\xi=1$ there survived some red points allowed by all constraints,
which implies that  produced  GWs may be detectable by SKA, THEIA, BBO, LISA, TianQin or DECIGO, as shown in  Fig. \ref{fig3}.  As the peak frequency increases,  the power $\Omega_{\rm GW}$ decreases.   In brief,  for a temperature ratio $\xi$ smaller than $0.1$ the GW is not detectable; the detectable GW has a peak frequency of ($10^{-6}\sim 10^{-3}$) Hz for the hidden sector physics. 

In summary, with all constraints we find that the allowed dark matter mass space is 
about $(0.01\sim 10)$ GeV and the dark scalar mass is $(4\times 10^{-4} \sim   3)$ GeV for an attractive Yukawa potential. When the temperature rate between hidden sector and visible sector is  $0.1<\xi<1 $,  the produced GWs may be detectable in the future measurements which may serve as a probe of the hidden sector.  

\section{Conclusion}\label{sec5}
In this work, we considered a hidden SIDM sector with a Dirac fermion as dark matter and a scalar to mediate self-interaction of dark matter. This hidden sector is highly decoupled from the visible sector and is colder than the visible sector.
From a generic quartic finite-temperature potential with a cubic term,
we studied the induced strong first-order phase transition and the resulted gravitational waves, considering the constraints from the self-interacting dark matter and  the relic density of dark matter. 
We found that the mass range of the dark scalar is about   
$ (4\times 10^{-4} \sim  3)$ GeV for an attractive Yukawa potential of  the self-interacting dark matter. For the dark matter particle, when the temperature ratio $\xi>0.1$, its mass range is about $(10~\rm MeV \sim 10 ~\rm GeV)$.  In the survived parameter space allowed by all constraints,   
the observability of the induced gravitational waves depends on the temperature  ratio $\xi$.
For $\xi <0.1$ the induced gravitational waves are not detectable, while for $0.1<\xi<1$ the gravitational waves with peak frequency of  
($10^{-6}\sim 10^{-3})$ Hz may be detectable in future projects.

\section*{Data availability statement}
No data associated in our work.
\section*{Acknowledgements}
This work was supported by the National
Natural Science Foundation of China (NNSFC) under grant Nos. 11775012, 11821505 and 12075300,
by Peng-Huan-Wu Theoretical Physics Innovation Center (12047503), and by the Key Research Program of the Chinese Academy of Sciences, grant No. XDPB15.



\end{document}